\documentclass[10pt,conference,fleqn]{IEEEtran}
\IEEEoverridecommandlockouts
\usepackage{cite}
\usepackage{amsmath,amssymb,amsfonts}
\usepackage{algorithmic}
\usepackage{graphicx}
\usepackage[table,xcdraw]{xcolor}
\usepackage{textcomp}
\usepackage{multirow}
\usepackage{multicol}
\usepackage{booktabs}
\usepackage{bigstrut}
\usepackage{verbatim}
\usepackage{array}
\usepackage{rotating}
\usepackage{xcolor}
\usepackage[a4paper, total={184mm,239mm}]{geometry}
\def\BibTeX{{\rm B\kern-.05em{\sc i\kern-.025em b}\kern-.08em
    T\kern-.1667em\lower.7ex\hbox{E}\kern-.125emX}}
\begin{document}

\title{An Efficient and Scalable Clocking Assignment Algorithm for Multi-Threaded Multi-Phase Single Flux Quantum Circuits \\
%\thanks{This work was supported by DIA fellowship from USC}
}

\author{\IEEEauthorblockN{Robert S. Aviles, Xi Li, Lei Lu, Zhaorui Ni, Peter A. Beerel}
\IEEEauthorblockA{\textit{Department of Electrical and Computer Engineering, University of Southern California}, Los Angeles, USA \\\{rsaviles, xli497, lulei, zhaoruin, pabeerel\}@usc.edu}}
%\and

%}

\maketitle

\begin{abstract}
A key distinguishing feature of single flux quantum (SFQ) circuits is that each logic gate is clocked. This feature forces the introduction of path-balancing flip-flops to ensure proper synchronization of inputs at each gate. This paper proposes a polynomial time complexity approximation algorithm for clocking assignments that minimizes the insertion of path balancing 
buffers for multi-threaded multi-phase clocking of SFQ circuits. Existing SFQ multi-phase clocking solutions have been shown to effectively reduce the number of required buffers inserted while maintaining high throughput, however, the associated clock assignment algorithms 
have exponential complexity and can have prohibitively long runtimes for large circuits, limiting the scalability of this approach. Our proposed algorithm is based on a linear program (LP) that 
leads to solutions that are experimentally on average within 5\% of the optimum and helps accelerate convergence towards the optimal integer linear program (ILP) based solution.  The improved LP and ILP runtimes permit multi-phase clocking schemes to scale to larger SFQ circuits than previous state of the art clocking assignment methods. We further extend the existing algorithm to support fanout sharing of the added buffers, saving, on average, an additional 10\% of the inserted DFFs. Compared to traditional full path balancing (FPB) methods across 10 benchmarks, our enhanced LP saves 79.9\%, 87.8\%, and 91.2\% of the inserted buffers for 2, 3, and 4 clock phases respectively. Finally, we extend this approach to the generation of circuits that completely mitigate potential hold-time violations at the cost of either adding on average less than 10\% more buffers (for designs with 3 or more clock phases) or, more generally, adding a clock phase and thereby reducing throughput. 

\end{abstract}


\section{Introduction}

After decades of consistent improvements in semiconductor fabrication, transistor sizes are soon expected to reach a practical minimum \cite{end_moore} and a performance ceiling for \textit{complementary metal-oxide-semiconductor} (CMOS) devices is approaching. This looming end to Moore's Law, coupled with increasing power densities and a practical limit on clock frequencies for CMOS devices, has driven the demand for emerging device technologies to provide energy-efficient performance improvements. Superconductive \textit{single-flux quantum} (SFQ) \cite{isvlsi2} devices have become a promising replacement for CMOS to provide low power exasale supercomputing \cite{Sergey2016}. 

The high potential of SFQ, however, has yet to be realized due to several design challenges. The underlying device technology dictates that all logic gates are clocked \cite{isvlsi4}, producing deep gate-level pipelines that must be carefully managed.  The simplest approach to managing data flow is \textit{full path balancing} (FPB) where numerous D Flip-Flops (DFFs) are inserted to match fanin path lengths to each node. The number of inserted DFFs for this method can far exceed the number of gates and substantially increase area and power consumption.

Alternatively, the multi-phase clocking approach \cite{ISVLSI,XI_thesis} significantly reduces path balancing DFFs and enables efficient multi-threaded operation, where clock frequency matches the achieved throughput. However, the problem is assumed to be NP-hard and relies on an integer linear program (ILP) based optimization problem whose runtime does not scale to large circuits.
To address this scalability issue we propose a reformulation of the multi-phase algorithm that offers scalability through near-optimal approximation as a linear program (LP) and improved convergence towards an optimal solution as an ILP. 

In addition to scalability challenges, SFQ circuits must contend with significant variations caused by process variation and environmental conditions that introduce significant delay uncertainties \cite{Rami8,Rami9}. Given the gate-level pipelined nature of these circuits, logic delays are relatively small, making the circuits highly vulnerable to hold violations. This is important because, as in CMOS, a single hold violation can make a chip inoperable and slowing down the clock does not help. To address this issue, motivated by traditional two-phase CMOS clocking solutions \cite{CarverMead}, we extend our approach to produce 100\% hold-safe circuits in which any hold violations can be resolved by adjusting the clocks' frequency and phase shifts \cite{isvlsi3}. All of our methodologies are functionally verified across 10 benchmark circuits and experimental results and the contributions of this work can be summarized as follows:

\begin{itemize}
    \item We propose a multi-threaded multi-phase clocking assignment algorithm to minimize buffer insertion in multi-threaded pipelined SFQ circuits that provides polynomial-time LP solutions that save an additional 10\% buffers by implementing fanout-sharing-aware minimization. Compared to traditional full path balancing, our formulation saves, on average, over 90\% of the added buffers using four clock phases and is less than 5\% away from the ILP-based optimum (when known). 
    \item We provide an algorithmic extension to enable the generation of circuits that are 100\% robust against hold time violations at the cost of increased area or decreased throughput. More precisely, the cost of this increased robustness is on average less than 10\% more buffers for designs with 3 or more clock phases or adding a clock phase. 
    \item The increased scalability of our algorithm is measured and quantified across our benchmark circuits. For example, we optimize a Shoup multiplier for homomorphic encryption \cite{FPGAShoup}\cite{ShoupAlgo} in 6 seconds whereas the state of the art (SOTA) ILP-based algorithm does not return a feasible solution within 50 minutes.
\end{itemize}

\section{Background}
\subsection{SFQ Devices}
SFQ devices have the theoretical potential for power reduction of three orders of magnitude compared to SOTA semiconductor technologies \cite{thesis5}. SFQ technology is based on materials that show zero electrical resistance and magnetic field expulsion when they are cooled below a characteristic critical temperature. When the environment temperature is 4.2 Kelvin, the transmission of pulses has a speed approaching the speed of light, using passive superconductive microstrip lines
\cite{isvlsi3}. A recent study presents a taped-out 4-bit processor in SFQ logic operating at
32GHz \cite{thesis8}. Despite the high frequency and the cooling overhead, the devices in SFQ circuits still consume significantly less power compared to CMOS devices \cite{thesis9}.

SFQ logic gates process input pulses by changing the state of current loops internal to the gate, an output pulse is produced for a $'1'$ state in response to an input clock pulse \cite{isvlsi4} and the internal state is reset.  In this way each SFQ logic gate is a sequential circuit element that creates a deep gate-level-pipelined circuit.

\subsection{SFQ Clocking Methods}

\begin{figure}[h]
\includegraphics[width=\columnwidth]{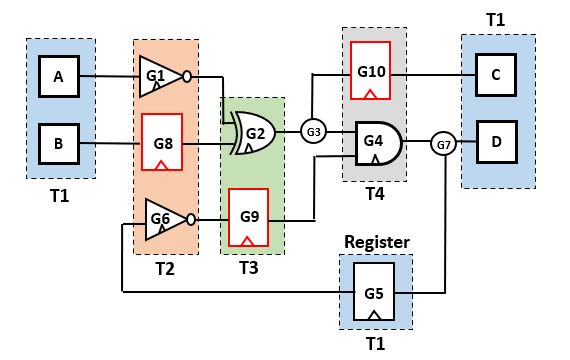}% This is a *.png file
%\vspace{-0.6cm}
\centering
\caption{Example Circuit Using Full Path Balancing \cite{ISVLSI}}
\label{fig:FPB}
\end{figure}

To ensure proper functionality, data flow through all resulting pipeline paths in a circuit must be properly synchronized at all points of reconvergence. The FPB method assumes all gates are driven by the same clock and inserts DFFs as needed to ensure that each path from a primary input to a given node takes the same number of clock cycles. For an 8-bit integer divider the inserted DFFs are 4.5x the original gate count \cite{Ghasem}.  Fig. \ref{fig:FPB} shows an example circuit using FPB clocking where the red buffers are inserted DFFs and shaded blocks indicate simultaneous threads of computation. Various methods have been proposed to decrease the inserted DFF cost.  

Dual Clocking Methods (DCM) \cite{Ghasem}\cite{Mingye} have been proposed where two clocks, one slow and one fast, have been used to remove path balancing DFFs at the cost of reducing throughput.  The faster clock is used to help propagate the data through the logic and notably must be at a higher frequency that the resulting throughput. Managing the two clocks complicates clock tree routing 
and timing closure issues. Importantly, this approach trades performance for area but does not mitigate 
the potential dangers of hold violations as it still requires a high speed clock.
%In our \textit{Experiments} section we compare the area/throughput benefits of DCM to our multi-phase clocking method. %is a partial path balancing method that

\begin{figure}[h]
\includegraphics[width=\columnwidth]{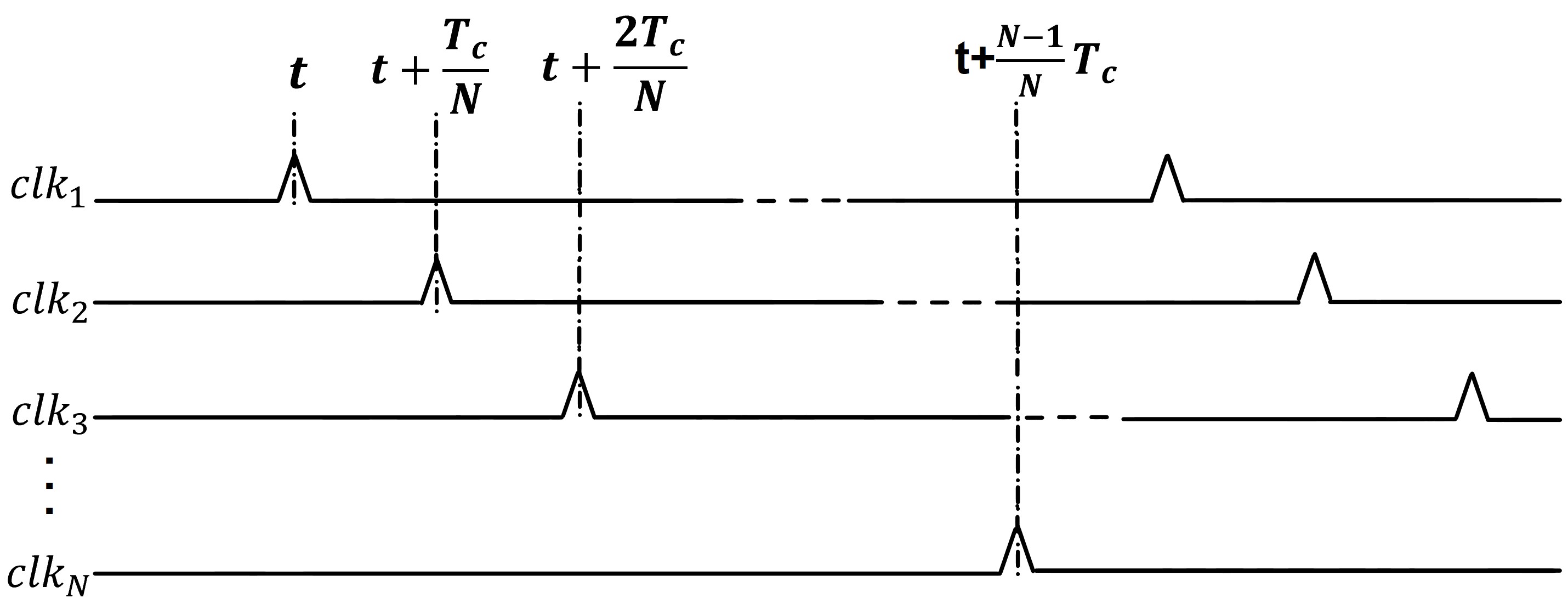}% This is a *.png file
%\vspace{-0.6cm}
\centering
\caption{Multi-Phase Clocks ($T_c$: cycle time)}
\label{fig:MP_clks}
\end{figure}

\begin{figure}[h]
\includegraphics[width=\columnwidth]{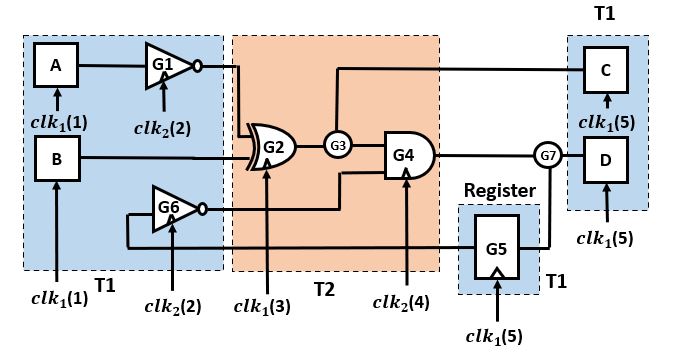}% This is a *.png file
%\vspace{-0.6cm}
\centering
\caption{Example Circuit Re-Designed with Two-Phase Clocking (depth label)}
\label{fig:MPCC}
\end{figure}

Alternatively, the multi-phase clocking method \cite{ISVLSI,XI_thesis} 
partitions the circuit into pipeline stages using multiple clocks with the same frequency but shifted phases as originally proposed for CMOS circuits in \cite{SMO} and as shown for SFQ clocks in Fig. \ref{fig:MP_clks}.   Optimal phase assignment and stage partitioning dramatically reduce the inserted DFF cost as clock phases can be utilized to balance paths instead of DFFs, Fig. \ref{fig:MPCC} shows the circuit of Fig. \ref{fig:FPB} re-designed using two-phase clocking to reduce the number of DFFs (in this example all DFFs were able to be removed). Since the maximum path delay between two circuit elements constrains the minimum period between arriving clock signals at the driving and receiving gate, the period between each clock phases $'1'$ pulse must remain fixed.  Accordingly, the clock period must increase with $N$. Notably, due to a proportional reduction in pipeline stages, the latency remains the same but the number of threads supported and maximum throughput scale by $\frac{1}{N}$. %Note that for $N$ clock phases, the number of threads supported is $\frac{1}{N}$ that of FPB.  %Generally, the maximum throughput of a multi-phase circuit will be $\frac{1}{N}$ FPB, but the latency will remain the same as FPB.  


\section{Linear Relaxation Compatible Algorithm}
\subsection{Key Insights}
The key simplification to our multi-phase clocking algorithm is the recognition that the assignment problem can be formulated based on a notion of the \textit{phase depth} (\textit{$D_i$}) of 
gate \textit{i}, which represents the total number of clock phases that span the propagation of data 
from the circuit inputs to gate \textit{i}.
This is in contrast to the SOTA which explicitly creates two optimization variables, pipeline stage $S_i$ and phase $CLK_i$ for each gate $i$ \cite{ISVLSI} in which each pipeline stage can support an independent thread of computation.  In Fig. \ref{fig:MPCC} we show the corresponding phase depth of each gate in parenthesis next to their clock phase. As an example, the sequence of phases required to propagate the inputs to output $C$ is ${clk_1,clk_2,clk_1,clk_2,clk_1}$, hence $C$ is given depth $5$ even though there are only four nodes between any input and $C$ (inclusive). Importantly, \textit{$D_i$} implicitly defines the gate's pipeline stage and clock phase. More precisely, $S_i = \lceil\frac{D_i}{N}\rceil$ and \textit{$CLK_i$ = $D_i$-$S_i$*N+N}.  It is important to emphasize that clock phase assignments are periodic with depth $N$, such that \textit{$CLK_j$=$CLK_i$} if \textit{$D_j$=$D_i$+N}.

Then two circuit elements $i$ and $j$ can be directly connected without DFF insertion if the depth difference between them is less than or equal to $N$, and a DFF is required to be inserted for all connections that exist with a phase depth difference of greater than $N$.  This bound on phase depth difference ensures that all inputs to gate $i$ have been processed on or after phase $CLK_i$ in $S_{i-1}$. This guarantees that any input to gate $i$ is ready and available to be processed on the $CLK_i$ phase of stage $S_i$.

\subsection{Proposed Algorithm}

With this phase depth constraint in mind we formulate the phase depth assignment as an optimization problem minimizing the total number of DFFs inserted. The circuit is modeled as a directed acyclic graph (DAG) where each gate is an internal node in the graph and circuit inputs (outputs) are input (output) nodes.  To model sequential circuits, all registers are inserted as a pair of psuedo input (PSI) and pseduo output (PSO) nodes.  A directed edge ($E_{ij}$) is created between each pair of nodes that is connected in the circuit (splitters and wires are abstracted as edges).  Each node is to be assigned a phase depth, which defines the number of DFFs that must be inserted between two connected gates.  For design interface consistency, all primary inputs (PI) are assigned phase depth $1$ and all primary outputs (PO) are forced to have equal phase depths. To ensure correct multi-threaded operation, all PSI/PSO pairs must have the same the phase depth difference between them, defined by a user-set  parameter $D_{loop}$, which forces all feedback loops to be the same length. $D_{loop}$ must be a multiple of $N$ so that a given register has the same clocking assignment for its PSI and PSO node abstractions and enables the circuit to support $\frac{D_{loop}}{N}$ concurrent threads.
The number of buffers to be inserted between connected nodes $i$ and $j$ is represented as $C_{ij}$.  The total number of buffers inserted is minimized subject to the constraints that phase depth is always increasing between connections, $D_j-D_i\geq{1}$, and that a DFF is inserted for each phase depth difference greater than $N$ that would exist in the resulting circuit, $D_j-D_i\leq{(C_{ij}+1)*N}$. The resulting ILP optimization problem is:
%\begin{comment}
\begin{equation} \label{eq:objective}
    \text{Minimize:} \sum_{E_{ij}=1}{C_{ij}}
\end{equation}
%\begin{align*} 
\begin{equation*}
    \text{subject to:}
\end{equation*}
\begin{equation}\label{eq:Depthdiff}
 1\leq D_j-D_i \leq {(C_{ij}+1)*N} \hspace*{0.3in}  \forall (i,j) \in E_{i,j},
\end{equation}
%\end{align*}
\begin{equation}
 D_i = D_{outputs}  \hspace*{0.3in} \forall i \in PO, 
\end{equation}
\begin{equation}
    D_i = 1    \hspace*{0.3in} \forall i \in PI,
\end{equation}
\begin{equation}
  D_j-D_k = D_{loop} \hspace*{0.3in}  \forall (j,k) \in (PSO,PSI),
\end{equation}
\begin{equation}  
\label{eq:Integral}
    D_i, C_{ij} \in \mathbb{N} \hspace*{0.3in}  \forall (i,j) \in E_{i,j}
\end{equation}
%\end{comment}

After a solution is returned, the assigned clock phases for each gate are extracted from the resulting $D_i$ assignments as described above and $C_{ij}$ DFFs are inserted into the circuit along the connection between gates $i$ and $j$, each with clock phase $CLK_i$.

\subsection{Linear Relaxation}\label{linrelax}

The key benefit of our formulation is that integer constraints in Eq. \ref{eq:Integral} can be relaxed and a feasible solution will still be found with minimal degradation of solution quality. In particular, 
to handle non-integer values, all returned \textit{$D_i$} values are rounded up to the nearest whole number. Since all values are rounded in the same direction, connected nodes that were subject to the relationship $D_j-D_i\geq{1}$ before rounding, will still satisfy the constraint after rounding. 

For $C_{ij}$ values, a post processing recalculation can be done where $C_{ij}=\lceil\frac{D_j-D_i}{N}-1\rceil$. This post processing is necessary to insert the minimum whole number of DFFs that satisfy the given depth assignments. After post processing and rounding for all of our $D_i$,$D_j$, and $C_{ij}$'s, constraints will be satisfied producing feasible circuits. 

Our optimization problem can thus be solved as a linear program (LP) with polynomial-time complexity. However, the linear relaxation does introduce some approximation into our solution.  Our LP will minimize the sum of fractional $C_{ij}$'s, when the true solution quality depends on the sum of whole number $C_{ij}$'s, causing our LP to incorrectly assess the cost associated with a solution.  However, this approximation cost tends to be very little as the exact cost for each $C_{ij}$ is the approximated cost ($C_{ij}=\frac{D_j-D_i}{N}-1$) rounded up.  For a lower number clock phases, as is seen commonly in practice, a small value of $N$ will introduce minimal approximation. 
\section{Hold Safe Circuits}\label{sect_holdsafe}

As discussed in \cite{XI_thesis}\cite{isvlsi3}, for a multi-phase clocked system whose time between clock phases can be adjusted after fabrication (notably increased) and where connected gates are clocked by different phases, there exists a minimum clock period for which the circuit can work correctly under any setup and hold conditions.\footnote{In particular, hold violations can be solved by increasing the distance between clock phases.} We refer to such circuits as {\em hold safe} and we present a reformulation of our optimization problem to satisfy these constraints. In particular, by subtracting $1$ from $N$ in our original formulation such that constraint (\ref{eq:Depthdiff}), becomes: 
\begin{center}
$1\leq{D_j-D_i}\leq{(C_{ij}+1)*(N-1)}$.\\
\end{center}  Recall that $D_j$ and $D_i$ are the connected nodes but there will be $C_{ij}$ total DFFs inserted between them.  

To clarify, we define $\Delta$$D_{ij}$ as the depth span of the connection between nodes $i$ and $j$ and $\Delta$$d_{mn}$ as the depth difference between any two circuit elements $m$ and $n$ along the path between $i$ and $j$.  The set of these circuit elements include nodes $i$, $j$ and the $C_{ij}$ DFFs to be inserted between them. A minimum $C_{ij}$ is then $1$ less than the number of segments of depth difference $N-1$ within span $\Delta$$D_{ij}$.  This enforces a DFF insertion for any $\Delta$$d_{mn}>N-1$ that would exist in the path, allowing $C_{ij}$ DFFs to be inserted such that all $\Delta$$d_{mn}$$\leq$$N-1$.  Since clock phase assignments are periodic with $N$ depth this ensures that no connected elements will have the same clock phase assignments.  When inserting DFFs for timing robust circuits we insert them for maximum span $\Delta$$d_{mn}=N-1$, each DFF is assigned to the clock phase preceding the clock of its fanin node.  This tighter constraint requires that for the same number of clock phases additional DFFs are needed.

Interestingly, as will be shown in our results, the hold safe formulation will a have nearly identical DFF cost as a non-robust solution with $1$ less clock phase, with the key difference that the boundary constraint of $D_{loop}$ is still forced to be a multiple of $N$ to ensure that pseudo inputs and outputs have the same clock phase assignment. This implies that the cost associated with a hold safe circuit is typically either the introduction of an additional clock phase with the same number of DFFs 
(reducing throughput) or maintaining the same number of clock phases and adding additional DFFs. 
%One difference between 3 clock phases hold safe and 2 clock phases non-robust will be that a solution value $D_i$ is still interpreted with respect to 3 clock phases and therefore will have a different clock assignment. 

\section{Fanout-Sharing Aware Minimization}
Our previous formulations have minimized the total number of DFFs being inserted between connections in a circuit.  However, DFFs can be optimally inserted as a linear pipeline between a driver and its fanout nodes, allowing some inserted DFFs to be shared amongst connections \cite{XI_thesis}.  The linear pipeline is constructed by inserting the largest number of DFFs between a driver and any of its fanouts.  Each fanout can then be connected to the DFF in the pipeline that would satisfy its particular connection requirement.  

To utilize this connection sharing, we reformulated our optimization method to minimize the total number of DFFs when inserted as a linear pipeline by introducing a new variable $C_i$ which is set to be the maximum of all $C_{ij}$ costs for each $i,j$ connection by adding the constraint $C_{ij}\leq{C_i}$. 
The objective function is then simplified to be the sum of $C_i$ to represent the number of added DFFs.

For each connection, $C_{ij}$ is no longer minimized but upper bounded by their respective $C_i$, 
we must recalculate $C_{ij}$ as described in Section \ref{linrelax} from the obtained values of 
$D_i$ and $D_j$ for both ILP and LP implementations to recover their correct values 
and associated connections.

\section{Experimental Results}
\subsection{Verification Environment}

All of our benchmark circuits were synthesized from an initial CMOS based Verilog file and our resulting circuits were tested through simulation using an SFQ emulating Verilog library as presented in \cite{Ramiverilog}. Test patterns consisted of 1,000 randomly generated input vectors per thread of operation. Combinational circuits were maximally pipelined, and sequential circuits were maximally threaded.  The resulting outputs were compared to golden results obtained from simulating the original Verilog file for the same input sequences.  For verification of multiple threads, golden results were obtained for each individual thread and compared to the extracted output of each thread of our simulated SFQ circuit.

\subsection{DFF Insertion Results}

Compared to FPB, our LP and ILP substantially reduced the cost of inserted DFFs across the 10 benchmarks circuits. Resulting savings were calculated as percentage of total DFFs saved across the 10 benchmark circuits. The ILP was permitted a runtime of 50 min and any timeout results that returned a suboptimal solution are notated *.  It should also be noted that unlike our multi-phase circuits, the FPB circuits were not implemented to support multi-threading.  Using our optimized fanout sharing aware algorithm (Table \ref{tab:DFFs}), our algorithm achieved savings of 79.9\%, 87.8\%, and 91.2\%  of the inserted buffers for the LP, while the optimal ILP saves 80.5\%, 88.3\%, and 91.6\%; for 2, 3, and 4 clock phases respectively. Compared to the baseline multi-phase clocking, the fanout aware implementation provides additional savings of 10.3\%, 5.6\%, and 3.6\% of inserted DFFs, for 2, 3, and 4 clock phases respectively.  Note that for both methods, a reduction in DFF count correlates to a reduction in clock tree splitters as well.
% Table generated by Excel2LaTeX from sheet 'Base'
% Table generated by Excel2LaTeX from sheet 'Base'
\begin{table*}[t]
  \centering
  \caption{Required DFFs vs FPB}
    \begin{tabular}{|p{1.25cm}|p{0.75cm}|p{0.75cm}|p{0.75cm}|p{0.75cm}|p{0.75cm}|p{0.75cm}|p{0.75cm}|p{0.75cm}|p{0.75cm}|p{0.75cm}|p{0.75cm}|p{0.75cm}|p{0.75cm}|}
    \hline
    \multicolumn{1}{|c|}{\multirow{4}[8]{*}{\textbf{Benchmark}}} &
      \multicolumn{13}{c|}{\textbf{Inserted DFFs}}
      \bigstrut\\
\cline{2-14}     &
      \multirow{3}[6]{*}{\textbf{FPB}} &
      \multicolumn{6}{c|}{\textbf{Multi-Phase Clocking}} &
      \multicolumn{6}{c|}{\textbf{Fanout Aware Multi-Phase Clocking}}
      \bigstrut\\
\cline{3-14}     &
       &
      \multicolumn{2}{c|}{\textbf{2-Phase}} &
      \multicolumn{2}{c|}{\textbf{3-Phase}} &
      \multicolumn{2}{c|}{\textbf{4-Phase}} &
      \multicolumn{2}{c|}{\textbf{2 -Phase}} &
      \multicolumn{2}{c|}{\textbf{3-Phase}} &
      \multicolumn{2}{c|}{\textbf{4-Phase}}
      \bigstrut\\
\cline{3-14}     &
       &
      \textbf{LP} &
      \textbf{ILP} &
      \textbf{LP} &
      \textbf{ILP} &
      \textbf{LP} &
      \textbf{ILP} &
      \textbf{LP} &
      \textbf{ILP} &
      \textbf{LP} &
      \textbf{ILP} &
      \textbf{LP} &
      \textbf{ILP}
      \bigstrut\\
    \hline
    \textbf{s27} &
      18 &
      6 &
      6 &
      4 &
      3 &
      1 &
      1 &
      6 &
      6 &
      4 &
      3 &
      1 &
      1
      \bigstrut\\
    \hline
    \textbf{s298} &
      133 &
      63 &
      62 &
      40 &
      36 &
      22 &
      22 &
      51 &
      50 &
      37 &
      33 &
      21 &
      21
      \bigstrut\\
    \hline
    \textbf{s382} &
      255 &
      91 &
      87 &
      54 &
      48 &
      27 &
      26 &
      80 &
      72 &
      52 &
      44 &
      27 &
      26
      \bigstrut\\
    \hline
    \textbf{s526} &
      234 &
      85 &
      80 &
      44 &
      42 &
      28 &
      28 &
      70 &
      65 &
      44 &
      41 &
      28 &
      28
      \bigstrut\\
    \hline
    \textbf{Counter64} &
      969 &
      781 &
      748 &
      454 &
      446 &
      323 &
      312 &
      595 &
      580 &
      368 &
      363 &
      281 &
      277
      \bigstrut\\
    \hline
    \textbf{c3540} &
      1,694 &
      625 &
      596 &
      310 &
      286 &
      201 &
      184 &
      473 &
      445 &
      246 &
      231 &
      169 &
      159
      \bigstrut\\
    \hline
    \textbf{c5315} &
      4,778 &
      2,132 &
      2,037 &
      1,202 &
      \textit{1,134*} &
      798 &
      729 &
      1,410 &
      1,368 &
      803 &
      767 &
      554 &
      522
      \bigstrut\\
    \hline
    \textbf{c7552} &
      3,393 &
      1,313 &
      1,256 &
      778 &
      \textit{748*} &
      421 &
      407 &
      1,065 &
      1,012 &
      649 &
      \textit{630*} &
      360 &
      350
      \bigstrut\\
    \hline
    \textbf{c6288} &
      9,524 &
      2,639 &
      2,563 &
      1,570 &
      \textit{1,494*} &
      1,075 &
      \textit{1,023*} &
      1,766 &
      1,727 &
      1,097 &
      \textit{1,054*} &
      797 &
      749
      \bigstrut\\
    \hline
    \textbf{Shoup\_MM} &
      41,118 &
      11,136 &
      \textit{10,838*} &
      6,600 &
      \textit{6,436*} &
      4,795 &
      \textit{4,609*} &
      6,984 &
      \textit{6,787*} &
      4,264 &
      \textit{4,128*} &
      3,219 &
      \textit{3,100*}
      \bigstrut\\
    \hline
    \textbf{Total DFFs} &
      62,116 &
      18,871 &
      18,273 &
      11,056 &
      10,673 &
      7,691 &
      7,341 &
      12,500 &
      12,112 &
      7,564 &
      7,294 &
      5,457 &
      5,233
      \bigstrut\\
    \hline
    \rowcolor[rgb]{ .906,  .902,  .902} \multicolumn{2}{|c|}{\textbf{Total Savings}} &
      \textbf{69.60\%} &
      \textbf{70.60\%} &
      \textbf{82.20\%} &
      \textbf{82.80\%} &
      \textbf{87.60\%} &
      \textbf{88.20\%} &
      \textbf{79.90\%} &
      \textbf{80.50\%} &
      \textbf{87.80\%} &
      \textbf{88.30\%} &
      \textbf{91.20\%} &
      \textbf{91.60\%}
      \bigstrut\\
    \hline
    \end{tabular}%
  \label{tab:DFFs}%
\end{table*}%

For our improved fanout-aware algorithm we also calculated the deviation from optimal of buffers inserted for our LP. Instead of buffers saved from FPB, this metric was calculated as the percentage of extra buffers inserted across the benchmarks compared to the optimal ILP solution. The calculated ILP to LP cost is 3.59\%, 4.85\%, and 4.90\% extra buffers for 2, 3 and 4 clock phases. 

Our hold safe algorithm (Table \ref{tab:holdlp}) used the same metrics as our baseline, but only for 3 and 4 clock phases, as 2 clock phases with hold safe constraints has a similar DFF cost as full path balancing.  As discussed in Section \ref{sect_holdsafe}, the DFF cost of hold safe circuits is nearly identical to the non-robust solution with $1$ less clock phase.  
The fanout-aware optimization can be applied to our hold safe method with the same expectation of requiring 1 extra clock phase to get similar DFF insertion results as the fanout-aware methods in Table \ref{tab:DFFs}.

% Table generated by Excel2LaTeX from sheet 'HoldSafe'
\begin{table}[htbp]
  \centering
  \caption{Required DFFs for Hold Safe Formulation vs FPB}
    \begin{tabular}{|c|c|c|c|c|c|}
    \hline
    \multirow{4}[8]{*}{\textbf{Benchmark}} &
      \multicolumn{5}{c|}{\textbf{Inserted DFFs}}
      \bigstrut\\
\cline{2-6}     &
      \multirow{3}[6]{*}{\textbf{FPB}} &
      \multicolumn{4}{c|}{\textbf{Multi-Phase Clocking}}
      \bigstrut\\
\cline{3-6}     &
       &
      \multicolumn{2}{c|}{\textbf{3-Phase}} &
      \multicolumn{2}{c|}{\textbf{4-Phase}}
      \bigstrut\\
\cline{3-6}     &
       &
      \textbf{LP} &
      \textbf{ILP} &
      \textbf{LP} &
      \textbf{ILP}
      \bigstrut\\
    \hline
    \textbf{s27} &
      18 &
      8 &
      8 &
      2 &
      2
      \bigstrut\\
    \hline
    \textbf{s298} &
      133 &
      87 &
      79 &
      40 &
      36
      \bigstrut\\
    \hline
    \textbf{s382} &
      255 &
      114 &
      105 &
      54 &
      48
      \bigstrut\\
    \hline
    \textbf{s526} &
      234 &
      113 &
      102 &
      44 &
      42
      \bigstrut\\
    \hline
    \textbf{Counter64} &
      969 &
      884 &
      851 &
      541 &
      526
      \bigstrut\\
    \hline
    \textbf{c3540} &
      1,694 &
      625 &
      596 &
      310 &
      286
      \bigstrut\\
    \hline
    \textbf{c5315} &
      4,778 &
      2,132 &
      2,037 &
      1,202 &
      \textit{1,134*}
      \bigstrut\\
    \hline
    \textbf{c7552} &
      3,393 &
      1,313 &
      1,256 &
      778 &
      \textit{748*}
      \bigstrut\\
    \hline
    \textbf{c6288} &
      9,524 &
      2,639 &
      2,563 &
      1,570 &
      \textit{1,494*}
      \bigstrut\\
    \hline
    \textbf{Shoup\_MM} &
      41,118 &
      11,136 &
      \textit{10,841*} &
      6,600 &
      \textit{6,436*}
      \bigstrut\\
    \hline
    \textbf{Total DFFs} &
      62,116 &
      19,051 &
      18,434 &
      11,141 &
      10,752
      \bigstrut\\
    \hline
    \rowcolor[rgb]{ .906,  .902,  .902} \multicolumn{2}{|c|}{\textbf{Total Savings }} &
      \textbf{69.3\%} &
      \textbf{70.3\%} &
      \textbf{82.1\%} &
      \textbf{82.7\%}
      \bigstrut\\
    \hline
    \end{tabular}%
  \label{tab:holdlp}%
\end{table}%

\subsection{Approximated Two-Phase versus Dual Clocking Solutions}

We compare returned LP solutions for multi-phase clocking with reported DCM results for combinational circuits from \cite{Ghasem}. Thoughputs are normalized with respect to FPB implementations, where our throughput will be $1/N$.  As can be seen in Table \ref{Comb DCM}, the no path balancing implementation (NPB) of DCM can remove all DFFs, however, the throughput is reduced dramatically.\footnote{In Tables \ref{Comb DCM} \& \ref{Sequential DCM}, T is an abbreviation for Throughput}    To improve throughput in DCM, partial path balancing (PPB) is used at the cost of inserting DFFs. Our work outperforms reported results for PPB in both throughput and inserted DFF count.

% Table generated by Excel2LaTeX from sheet 'Combinational compare'
\begin{table}[htbp]
  \centering
  \caption{DCM vs Two-Phase for Combinational Circuits}
    \begin{tabular}{|c|c|c|c|c|c|c|}
    \hline
    \multicolumn{1}{|c|}{\multirow{4}[6]{*}{\textbf{Benchmark }}} &
      \multicolumn{6}{c|}{\textbf{Insertion Method}}
      \bigstrut\\
\cline{2-7}     &
      \multicolumn{2}{c|}{\multirow{2}[2]{*}{\textbf{DCM (NPB)}}} &
      \multicolumn{2}{c|}{\multirow{2}[2]{*}{\textbf{DCM (PPB)}}} &
      \multicolumn{2}{c|}{\multirow{2}[2]{*}{\textbf{Two-Phase}}}
      \bigstrut[t]\\
     &
      \multicolumn{2}{c|}{} &
      \multicolumn{2}{c|}{} &
      \multicolumn{2}{c|}{}
      \bigstrut[b]\\
\cline{2-7}     &
      \textbf{T} &
      \textbf{\#DFFs} &
      \textbf{T} &
      \textbf{\#DFFs} &
      \textbf{T} &
      \textbf{\#DFFs}
      \bigstrut\\
    \hline
    \textbf{c3540} &
      1/32 &
      0 &
       1/5 &
      795 &
       1/2 &
      473
      \bigstrut\\
    \hline
    \textbf{c5315} &
       1/27 &
      0 &
       1/5 &
      2554 &
       1/2 &
      1410
      \bigstrut\\
    \hline
    \textbf{c7552} &
       1/21 &
      0 &
       1/5 &
      1912 &
       1/2 &
      1065
      \bigstrut\\
    \hline
    \end{tabular}%
  \label{Comb DCM}%
\end{table}%

% Table generated by Excel2LaTeX from sheet 'Sequential compare'
\begin{table}[htbp]
  \centering
  \caption{DCM vs Two-Phase for Sequential Circuits}
    \begin{tabular}{|c|c|c|c|c|c|c|}
    \hline
    \multicolumn{1}{|c|}{\multirow{4}[6]{*}{\textbf{Benchmark }}} &
      \multicolumn{6}{c|}{\textbf{Insertion Method}}
      \bigstrut\\
\cline{2-7}     &
      \multicolumn{2}{c|}{\multirow{2}[2]{*}{\textbf{DCM (LTDC)}}} &
      \multicolumn{2}{c|}{\multirow{2}[2]{*}{\textbf{DCM (HTDC)}}} &
      \multicolumn{2}{c|}{\multirow{2}[2]{*}{\textbf{Two-Phase }}}
      \bigstrut[t]\\
     &
      \multicolumn{2}{c|}{} &
      \multicolumn{2}{c|}{} &
      \multicolumn{2}{c|}{}
      \bigstrut[b]\\
\cline{2-7}     &
      \textbf{T} &
      \textbf{\#Cells} &
      \textbf{T} &
      \textbf{\#Cells} &
      \textbf{T} &
      \textbf{\#Cells}
      \bigstrut\\
    \hline
    \textbf{s27} &
      1/9 &
      25 &
      1 &
      131 &
      1/2 &
      23
      \bigstrut\\
    \hline
    \textbf{s298} &
      1/12 &
      145 &
      1 &
      450 &
      1/2 &
      173
      \bigstrut\\
    \hline
    \textbf{s382} &
      1/13 &
      202 &
      1 &
      646 &
      1/2 &
      252
      \bigstrut\\
    \hline
    \textbf{s526} &
      1/12 &
      256 &
      1 &
      658 &
      1/2 &
      296
      \bigstrut\\
    \hline
    \textbf{Counter64} &
      1/15 &
      888 &
      1 &
      3,379 &
      1/2 &
      1,352
      \bigstrut\\
    \hline
    \end{tabular}%
  \label{Sequential DCM}%
\end{table}%

For sequential circuits we compare our work to DCM results obtained using calculations for total cell count and throughput from \cite{Mingye}. In Table \ref{Sequential DCM} the low throughput dual clocking (LTDC) implementation prioritizes minimizing cell count at the cost of throughput.  The high throughput dual clocking (HTDC) implementation was introduced, increasing the throughput with a high cell count overhead.  Despite providing a much higher throughput, our work is comparable to the LTDC in terms of cell count (1.2x average).  The HTDC can provide twice the throughput of our work but with a cell overhead that exceeds the circuit size of FPB implementations.  

Without considering timing complexity and potential yield issues, a DCM implementation may be preferred in cases where area for combinational circuits is substantially more important than the throughput and where throughput of sequential circuits can be prioritized at a steep area cost.  However, our work offers competitive benefits over DCM for many other situations and offers a balance of high throughput and low area for both sequential and combinational circuits. It should also be emphasized that DCMs requires the routing of two clocks, one with high frequency. This means DCM suffers from the same high risk of hold violations as the full path balancing single clock solution. Additionally, all DCM implementations repeat input pulses multiple times increasing energy costs. 

\subsection{Improved Runtime Scaling}
To demonstrate the scalability of our approximated LP solution over optimal solutions, we report in Table \ref{tab:time} the runtimes for the clocking assignment step of synthesizing the benchmark circuits with 2 clock phases.  The results also illustrate the utility of a feasible relaxed solution in speeding up the ILP's search for an optimal solution compared to the current SOTA.  With our algorithm the ILP can quickly find a feasible, near optimal solution for even very large circuits and then can be left to improve the solution quality according to the designer's need.  In contrast, the current SOTA \cite{ISVLSI} is not guaranteed to produce a feasible solution in a reasonable amount of time as a feasible solution for a single Shoup modular multiplier cannot be reached over a 50 min runtime.    
% Table generated by Excel2LaTeX from sheet 'Runtimes'
\begin{table}[htbp]
  \centering
  \caption{Runtimes for our methods vs SOTA}
    \begin{tabular}{|c|c|c|c|}
    \hline
    \multirow{2}[4]{*}{\textbf{Benchmark}} &
      \multicolumn{3}{c|}{\textbf{Runtimes for 2-Phase Clocking}}
      \bigstrut\\
\cline{2-4}     &
      \textbf{LP} &
      \textbf{ILP} &
      \textbf{Current SOTA}
      \bigstrut\\
    \hline
    \textbf{s27} &
      0s &
      0s &
      0s
      \bigstrut\\
    \hline
    \textbf{s298} &
      0s &
      0s &
      0s
      \bigstrut\\
    \hline
    \textbf{s382} &
      0s &
      0s &
      0s
      \bigstrut\\
    \hline
    \textbf{s526} &
      0s &
      0s &
      0s
      \bigstrut\\
    \hline
    \textbf{Counter64} &
      0s &
      1s &
      0s
      \bigstrut\\
    \hline
    \textbf{c3540} &
      0s &
      9s &
      142s
      \bigstrut\\
    \hline
    \textbf{c5315} &
      0s &
      34s &
      77s
      \bigstrut\\
    \hline
    \textbf{c7552} &
      0s &
      9s &
      65s
      \bigstrut\\
    \hline
    \textbf{c6288} &
      1s &
      17s &
      139s
      \bigstrut\\
    \hline
    \textbf{Shoup\_MM} &
      6s &
      (Suboptimal) 50 min &
      (No Soln) 50 min
      \bigstrut\\
    \hline
    \end{tabular}%
  \label{tab:time}%
\end{table}%

\section{Conclusions}

As SFQ technology continues to advance and designs include larger and more complex circuits, methods to minimize inserted DFFs and their associated clock tree splitters will increase in importance. Our work improves upon, not just the runtime scalability, but also the solution quality of the SOTA as a result of our fanout-aware optimization, which for 2 clock phases saves an additional 10\% compared to standard multi-phase clocking. Our method provides a DFF minimization algorithm that achieves savings upwards of 79\% with a scalable runtime that will be able to support the implementation of multi-phase clocking methodologies on larger circuits.  Additionally, we provide designers the ability to produce 100\% hold safe circuits at the cost of one additional clock phase.  This can save significant engineering effort in complex timing validation and also increase yield in expensive superconducting chip fabrication. 

Further work can be done to adapt this methodology to optimal clock phase assignments in AQFP circuits \cite{Nphaseclk}. Additionally, clock skew scheduling \cite{UsefulSkew} may be explored for multi-phase clocking  to reduce the clock tree overhead.

%\endinput

\bibliographystyle{IEEEtran}
\bibliography{bibliography}

\begin{thebibliography}{10}
\providecommand{\url}[1]{#1}
\csname url@samestyle\endcsname
\providecommand{\newblock}{\relax}
\providecommand{\bibinfo}[2]{#2}
\providecommand{\BIBentrySTDinterwordspacing}{\spaceskip=0pt\relax}
\providecommand{\BIBentryALTinterwordstretchfactor}{4}
\providecommand{\BIBentryALTinterwordspacing}{\spaceskip=\fontdimen2\font plus
\BIBentryALTinterwordstretchfactor\fontdimen3\font minus
  \fontdimen4\font\relax}
\providecommand{\BIBforeignlanguage}[2]{{%
\expandafter\ifx\csname l@#1\endcsname\relax
\typeout{** WARNING: IEEEtran.bst: No hyphenation pattern has been}%
\typeout{** loaded for the language `#1'. Using the pattern for}%
\typeout{** the default language instead.}%
\else
\language=\csname l@#1\endcsname
\fi
#2}}
\providecommand{\BIBdecl}{\relax}
\BIBdecl

\bibitem{end_moore}
T.~N. Theis and H.-S. P.~Wong, ``The end of moore’s law: A new beginning for
  information technology,'' \emph{Computing in Science \& Engineering},
  vol.~19, no.~2, pp. 41--50, 2017.

\bibitem{isvlsi2}
K.~Likharev and V.~Semenov, ``{RSFQ} logic/memory family: a new
  josephson-junction technology for sub-terahertz-clock-frequency digital
  systems,'' \emph{IEEE {TASC}}, vol.~1, no.~1, pp. 3--28, 1991.

\bibitem{Sergey2016}
S.~Tolpygo, ``Superconductor digital electronics: Scalability and energy
  efficiency issues (review article),'' \emph{Low Temperature Physics},
  vol.~42, pp. 361--379, 05 2016.

\bibitem{isvlsi4}
D.~K. Brock, E.~K. Track, and J.~M. Rowell, ``Superconductor {IC}s: the
  100-{GHz} second generation,'' \emph{IEEE Spectrum}, vol.~37, pp. 40--46,
  2000.

\bibitem{ISVLSI}
X.~Li, M.~Pan, T.~Liu, and P.~A. Beerel, ``Multi-phase clocking for
  multi-threaded gate-level-pipelined superconductive logic,'' in \emph{2022
  IEEE Computer Society Annual Symposium on VLSI (ISVLSI)}, 2022, pp. 62--67.

\bibitem{XI_thesis}
X.~Li, ``Multi-phase clocking and hold time fixing for single flux quantum
  circuits,'' Ph.D. dissertation, University of Southern California, 2022.

\bibitem{Rami8}
P.~Bunyk, K.~Likharev, and D.~Zinoviev, ``{RSFQ} technology: Physics and
  devices,'' \emph{International Journal of High Speed Electronics and
  Systems}, vol.~11, 03 2001.

\bibitem{Rami9}
I.~Vernik, Q.~Herr, K.~Gaij, and M.~Feldman, ``Experimental investigation of
  local timing parameter variations in {RSFQ} circuits,'' \emph{IEEE {TASC}},
  vol.~9, no.~2, pp. 4341--4344, 1999.

\bibitem{CarverMead}
C.~Mead and L.~Conway, \emph{Introduction to VLSI Systems}.\hskip 1em plus
  0.5em minus 0.4em\relax Addison-Wesley, 1980.

\bibitem{isvlsi3}
K.~Gaj, E.~G. Friedman, and M.~J. Feldman, ``Timing of multi-gigahertz rapid
  single flux quantum digital circuits,'' \emph{Journal of VLSI signal
  processing systems for signal, image and video technology}, vol.~16, pp.
  247--276, 1997.

\bibitem{FPGAShoup}
S.~Kim, K.~Lee, W.~Cho, J.~H. Cheon, and R.~A. Rutenbar, ``{FPGA}-based
  accelerators of fully pipelined modular multipliers for homomorphic
  encryption,'' in \emph{2019 International Conference on ReConFigurable
  Computing and FPGAs (ReConFig)}, 2019, pp. 1--8.

\bibitem{ShoupAlgo}
V.~Shoup, ``{NTL}: A library for doing number theory,''
  \url{http://www.shoup.net/ntl/}.

\bibitem{thesis5}
O.~A. Mukhanov, ``Energy-efficient single flux quantum technology,'' \emph{IEEE
  {TASC}}, vol.~21, no.~3, pp. 760--769, 2011.

\bibitem{thesis8}
K.~Ishida, M.~Tanaka, I.~Nagaoka, T.~Ono, S.~Kawakami, T.~Tanimoto,
  A.~Fujimaki, and K.~Inoue, ``32 {GH}z 6.5 m{W} gate-level-pipelined 4-bit
  processor using superconductor single-flux-quantum logic,'' in \emph{2020
  IEEE Symposium on VLSI Circuits}, 2020, pp. 1--2.

\bibitem{thesis9}
Q.~Herr, A.~Y. Herr, O.~T. Oberg, and A.~Ioannidis, ``Ultra-low-power
  superconductor logic,'' \emph{Journal of Applied Physics}, vol. 109, p.
  103903, 2011.

\bibitem{Ghasem}
G.~Pasandi and M.~Pedram, ``An efficient pipelined architecture for
  superconducting single flux quantum logic circuits utilizing dual clocks,''
  \emph{IEEE {TASC}}, vol.~30, no.~2, pp. 1--12, 2020.

\bibitem{Mingye}
M.~Li, B.~Zhang, and M.~Pedram, ``Striking a good balance between area and
  throughput of {RSFQ} circuits containing feedback loops,'' \emph{IEEE
  T{ASC}}, vol.~33, no.~5, pp. 1--6, 2023.

\bibitem{SMO}
K.~A. Sakallah, T.~N. Mudge, and O.~A. Olukotun, ``{Optimal Clocking of
  Synchronous Systems},'' in \emph{TAU Workshop}, 1990, pp. 1--21.

\bibitem{Ramiverilog}
R.~N. Tadros, A.~Fayyazi, M.~Pedram, and P.~A. Beerel, ``System{V}erilog
  modeling of {SFQ} and {AQFP} circuits,'' \emph{IEEE {TASC}}, vol.~30, no.~2,
  pp. 1--13, 2020.

\bibitem{Nphaseclk}
R.~Saito, C.~L. Ayala, and N.~Yoshikawa, ``Buffer reduction via {N}-phase
  clocking in adiabatic quantum-flux-parametron benchmark circuits,''
  \emph{IEEE {TASC}}, vol.~31, no.~6, pp. 1--8, 2021.

\bibitem{UsefulSkew}
R.~Bairamkulov, T.~Jabbari, and E.~G. Friedman, ``Qu{CTS}—single-flux quantum
  clock tree synthesis,'' \emph{IEEE Transactions on Computer-Aided Design of
  Integrated Circuits and Systems}, vol.~41, no.~10, pp. 3346--3358, 2022.

\end{thebibliography}

%%%%% USE THIS FOR BLIND REVIEW IN aaa_main.bib %%%%%
%\begin{comment}
%@electronic{usc-dnn-rtl,
% title     = "Removed for blind review"
%}
%\end{comment}

%%%%% USE THIS FOR NON-BLIND REVIEW IN aaa_main.bib %%%%%

\end{document}